\documentclass[runningheads]{llncs}
\usepackage{graphicx}
\usepackage{amssymb}
\usepackage{multirow}
\usepackage{booktabs}
\usepackage[misc]{ifsym}
\usepackage{graphicx}
\usepackage{amsmath}
\usepackage{booktabs}
\usepackage{algorithmic}
\usepackage{amssymb}
\usepackage{bm}
\usepackage{array}
\usepackage[T1]{fontenc}
\usepackage{epsfig}
\usepackage{psfrag}
\usepackage{subfigure}
\usepackage{epstopdf}
\usepackage{epsfig}
\usepackage{color}
\usepackage[ruled]{algorithm2e}
\usepackage{color}

\usepackage{CJKutf8}
\AtBeginDocument{\begin{CJK}{UTF8}{gbsn}}
	\AtEndDocument{\end{CJK}}
\begin{document}
\title{Schizophrenia detection based on EEG using Recurrent Auto-Encoder framework}
\titlerunning{Schizophrenia detection using Recurrent Auto-Encoder framework}
%
\author{Yihan Wu\inst{1} \and
Min Xia\inst{1} \and
Xiuzhu Wang\inst{1} \and
Yangsong Zhang\inst{1,2}\textsuperscript{(\Letter)}
}
\authorrunning{Y. Wu et al.}
%

\institute{School of Computer Science and Technology, Laboratory for Brain Science and Medical Artificial Intelligence, Southwest University of Science and Technology, Mianyang  621010 China \and
MOE Key Lab for Neuroinformation, University of Electronic Science and Technology of China, Chengdu 610054, China.\\
\email{zhangysacademy@gmail.com}
}
\maketitle              
\begin{abstract}
Schizophrenia (SZ) is a serious mental disorder that could seriously affect the patient's quality of life. In recent years, detection of SZ based on deep learning (DL) using electroencephalogram (EEG) has received increasing attention. In this paper, we proposed an end-to-end recurrent auto-encoder (RAE) model to detect SZ. In the RAE model, the raw data was input into one auto-encoder block, and the reconstructed data were recurrently input into the same block. The extracted code by auto-encoder block was simultaneously served as an input of a classifier block to discriminate SZ patients from healthy controls (HC).  Evaluated on the dataset containing 14 SZ patients and 14 HC subjects, and the proposed method achieved an average classification accuracy of 81.81\% in subject-independent experiment scenario. This study demonstrated that the structure of RAE is able to capture the differential features between SZ patients and HC subjects.

\end{abstract}

\section{Introduction}

Schizophrenia is a severe mental disorder. This disease affects approximately 24 million people in the world, reported by the World Health Organization~\cite{WHO}. One in 300, on average, people suffer from SZ, and this rate reaches up to one in 222 in adults~\cite{Rate}. However, the majority of patients with SZ have not received proper treatment. One of the most difficult issues is the absence of significant biological markers~\cite{9257371}.

Benefiting from the advantages such as non-invasive, high temporal resolution, low cost, electroencephalography (EEG) has been widely used in the disease detection field~\cite{8681406,saeedi2021major,BOOSTANI20096492,ACHARYA2018103}. With the development of machine learning, artificial features based on EEG signals have been rapidly employed in the field of SZ detection. For example, Vázquez et al. proposed a method using random forest to operate on the extracted connectivity metrics of generalized partial directed coherence (GPDC) and direct directed transfer function (dDTF) of 1-minute segments. They conduct subject-unaware partitioning and leave-$p$-subject-out experiments and obtain the area under the curve (AUC) of 0.99 and 0.87, respectively~\cite{vazquez2021interpretable}. Najafzadeh et al. proposed a method based on the adaptive neuro fuzzy inference system (ANFIS). They tried to employ ANFIS, support vector machine (SVM), and artificial neural network (ANN) to detect the SZ using Shannon entropy, spectral entropy, approximate entropy, and the absolute value of the highest slope of auto-regressive coefficients and achieved accuracy of 99.92\% in the subject-dependent experiment~\cite{najafzadeh2021automatic}. Chandran et al. introduced their method based on Long Short-Term Memory (LSTM). They calculated Katz fractal dimension, approximate entropy and the time-domain feature of variance as artificial feature, and fed them into the LSTM network to distinguish the SZ patients from HC subject. They obtained an accuracy of 99.0\% in the subject-dependent experiment~\cite{10.1007/978-981-15-5243-4_19}.

These methods utilized artificial features that are highly dependent on the prior knowledge of researchers. The outstanding high performance of deep learning makes end-to-end SZ detection possible. For instance, the CNN-LSTM model is proposed by Shoeibi et al. ​They tried several combinations of 1D-CNN and LSTM to verify the best model. Their model achieved an accuracy of 99.25\% in the subject-dependent experiment~\cite{shoeibi2021automatic}. Oh et al. introduced a deep convolution neural network (CNN) to detect SZ. This model contains four convolution layers, five max-pooling layers and two fully connected layers. The experiments were conducted in both subject-dependent and subject-independent scenarios using 25 second segments. They achieved an accuracy of 98.07\% and 81.26\% respectively~\cite{oh2019deep}.

In most of the studies presented, the methods were evaluated in a subject-dependent scenario, which has a serious problem called data leakage. Due to the high correlation between continuous EEG signals, when the EEG signals collected in one subject were divided into several segments, and these segments were shuffled and partitioned simultaneously into training set and testing set. The training set and the testing set were inevitably intersecting. On the other hand, logically speaking, the subject-dependent method is unpractical, as it is unreasonable to detect SZ for subjects after knowing clearly whether they are patients or not.

Based on this consideration, we proposed a model named Recurrent Auto-Encoder (RAE), and evaluated its performance in a subject-independent scenario. It contains a recurrent auto-encoder to extract task-related features and a linear classifier to recognize the SZ and HC. We conducted experiments on a publicly accessed dataset containing 14 schizophrenia patients and 14 age-matched healthy control subjects. The results indicate that our RAE performed better than the current baseline methods.

This paper is organized as follows. Section 2 introduces the dataset and proposed model. Section 3 describes the experiment setting and result. In Section 4, the discussions and conclusions are present.

\section{Materials and Methods}

\subsection{Dataset}
In this study, we used a dataset collected by the Institute of Psychiatry and Neurology in Warsaw, Poland~\cite{olejarczyk2017graph}. This dataset consists of EEG recording from 14 patients (7 males: 27.9 $\pm$ 3.3 years, 7 females: 28.3 $\pm$ 4.1 years) with SZ and 14 HC  (7 males: 26.8 $\pm$ 2.9, 7 females: 28.7 $\pm$ 3.4 years). The eyes-closed resting state EEG signals lasting for 15 minutes were collected with a sampling rate of 250 Hz. The 19 electrodes were used, i.e., Fp1, Fp2, F7, F3, Fz, F4, F8, T3, C3, Cz, C4, T4, T5, P3, Pz, P4, T6, O1 and O2, which were placed according to the standard of international 10-20 system. More details could be found in the reference~\cite{olejarczyk2017graph}.

\subsection{Pre-processing}
​To improve the signal-noise ratio, we first employed a bandpass filter with a frequency of 0.5-50 Hz. The data were then divided into segments of 5 s in length. The obtained segments should pass a threshold check to reduce the interference of electrooculography (EOG). We dropped the segment which peak value is out of range of $-100 \mu V \sim100 \mu V$. Finally, the common reference and z-score normalization were applied to obtain the processed data.

\subsection{Methods}


The motivation of the proposed model is that: if the EEG data are recurrently processed by a encoder-decoder is beneficial to generate more discriminative embedding codes, the procedure is summarized as follows:
\begin{itemize}
	\item [$\bullet$] Encode the data $\bm{D_1}$ to obtain the embedding $\bm{Z_1}$
	\item [$\bullet$] Decode the $\bm{Z_1}$ to reconstruct $\bm{D_2}$
	\item [$\bullet$] Process the $\bm{D_2}$ as above did for several loops to obtain $\bm{D_n}$ and $\bm{Z_n}$.
\end{itemize}

On the assumption that the encoder and decoder are effective and stable enough, the embedding codes $\bm{Z_1}$, ..., $\bm{Z_n}$ should remain similar task-related property, although the waveform of $\bm{D_1}$, ..., $\bm{D_n}$ maybe not exactly the same. We termed the similarity as semantic invariance. On the other hand, if we optimize the encoder to improve the semantic invariance between $\bm{Z_1}$, ..., $\bm{Z_n}$, the optimization could be regarded as effective. In actual application, the true label can be defined as the task-related property. Improving the prediction accuracy of all embedding codes, especially $\bm{Z_2}$,... ,$\bm{Z_n}$, can be regarded as improving semantic invariance. This is the key idea of this method.

Previous studies in the field of computer vision (CV) have proved that Auto-Encoder is a powerful frame of feature extraction and reconstruction~\cite{He_2022_CVPR}. Therefore, we leveraged the Auto-Encoder as the main architecture to design our model. EEGNet is a widely used baseline method in the field of EEG analysis~\cite{Lawhern_2018}. It has stable performance and feature representation ability. We designed the encoder and the corresponding decoder modules using the similar operations in EEGNet.

\begin{figure}[!htb]
	\centering
	\includegraphics[scale=0.25]{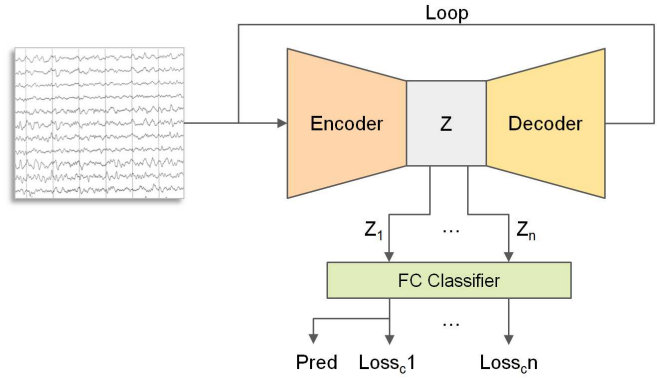}
	\caption{The structure of RAE. $Z_n$ is the representation recurrently generated by the encoder in cycles for $n$ times.}
	\label{Structure}
\end{figure}

The structure of RAE is shown in Fig.~\ref{Structure}. It is consisted of a recurrent auto-encoder feature extractor and a fully-connected classifier. The fully-connected classifier is used to classify all embedding codes extracted by RAE. The semantic invariance is improved by optimizing the classification accuracy to improve the performance of encoder.

\subsubsection{Backbone}

\begin{figure}[!htb]
	\centering
	\includegraphics[scale=0.30]{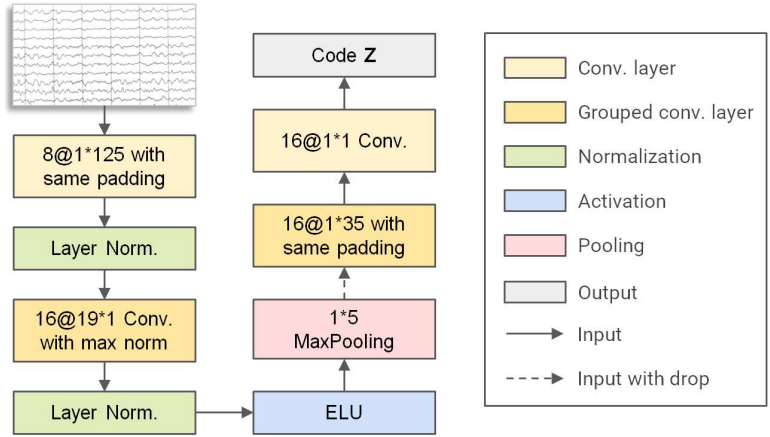}
	\caption{The structure of the encoder block. This block is denoted as $En$ in the formula.}
	\label{Backbone}
\end{figure}

The Backbone of RAE structure is modifiable. In this work, the backbones of the encoder and decoder were comprised of the similar operation that used in the classical EEGNet model. To facilitate decoding, the sizes of all temporal convolution kernels were set to be odd so that the padding can be symmetric. For similar reasons, the average pooling after the second convolution layer was replaced by a max pooling layer. ​In addition, we used layer normalization in the model in order to reduce the interference of other samples in one mini-batch. The structure is shown in Fig.~\ref{Backbone}

\begin{figure}[!htb]
	\centering
	\includegraphics[scale=0.305]{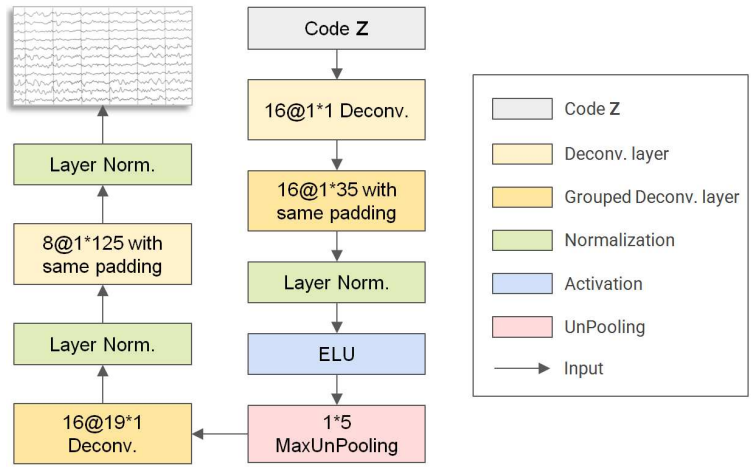}
	\caption{The structure of the decoder block. This block is denoted as $De$ in the formula. }
	\label{Decoder}
\end{figure}

​Decoder is the opposite procedure of encoder, which uses transposed convolution to realize deconvolution. In addition, layer normalization is applied in the end to keep each reconstructed sample separate from the others in one mini-batch. The structure of the decoder is shown in Fig.~\ref{Decoder}.

\subsubsection{Recurrent Auto-encoder}
First, the raw data $D_i\in\mathbb{R}^{C \times T}$ is input into the encoder block $En$ to generate the embedding code $Z_i$, which could be described as:
\begin{equation}
	Z_i =En(D_i) \in\mathbb{R}^{N*C'*T'}
\end{equation}
where $C'$ and $T'$ are the numbers of the electrode channels and time-dimension sampling point, which are equal to 1 and 250, respectively. $N$ denotes the number of the convolution kernels, which was set to 16 in this model.

The embedding code $Z_i$ is input into the decoder $De$ to reconstruct the data $D_{i+1}\in\mathbb{R}^{C \times T}$, which is illustrated in the following:
\begin{equation}
	D_{i+1} =De(Z_i) \in\mathbb{R}^{C*T}
\end{equation}

Then, the reconstructed data $D_{i+1}\in\mathbb{R}^{C \times T}$ was regarded as the input of block $En$ in the next cycle. After $n$ loop iterations, the model is able to generate embedding code $Z_1$,$Z_2$,..., $Z_n$. For the task of SZ detection, the $n$ was set to 2 in the following experiments. All the embedding codes will be employed to calculate the loss and predict the class label as follows.

\subsubsection{Classifier and Loss}
\begin{figure}[!htb]
	\centering
	\includegraphics[scale=0.3]{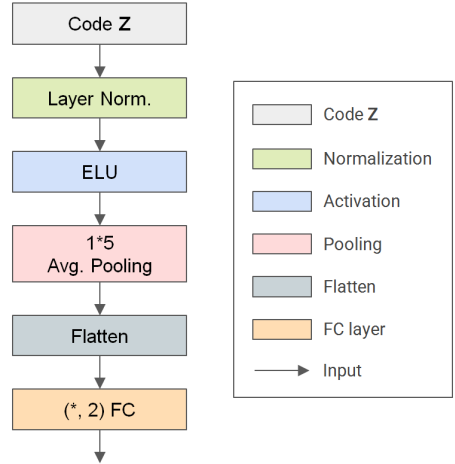}
	\caption{The structure of the classifier block. This block is denoted as $Cls$ in the formula.}
	\label{Classifier}
\end{figure}

The classifier structure is shown in the Fig.~\ref{Classifier}.
Embedding code $Z_i$ is input into the classifier $Cls$ to obtain the predicted label $\hat{y}_i$, which could be described as:
\begin{equation}
	{\hat{y}_i = \mathop{\arg\max} Cls(Z_i) } \label{cls}
\end{equation}

In this work, we employed the cross entropy loss between predicted labels and the corresponding true labels of the samples to optimize the model, which could be illustrated as:
\begin{equation}
	L_{i} = loss\_fn(\hat{y}_i, y)
\end{equation}
where $loss\_fn$ denotes cross entropy operator, and $y$ denotes true label.

\subsubsection{ Model training}
The parameters optimization of $En$, $De$ and $Cls$ blocks in each optimization loop were separate. The procedure is summarized as Algorithm \ref{training}:

\begin{center}
	\begin{minipage}{8cm}
		\begin{algorithm}[H]
			\caption{Training procedure}
			\label{training}
			\LinesNumbered 
			\KwIn{input parameters X, Y, nloops}
			\KwOut{Classifier Cls($\cdot$)}
			$X_0$ = X\;
			\For{i in range(0, nloops)}{
				$z_i$ = $En$($X_i$)\;
				$\hat{Y_i}$ = $Cls$($z_i$)\;
				$loss_i$ = LossFn($\hat{Y}$, Y)\;
				$loss_i\gets$ backward\;
				$X_{i+1}$ = $De$($z_i$)\;
			}
		\end{algorithm}
	\end{minipage}
\end{center}

In the testing stage, the for loop and code in lines 5,6 and 7 are unessential. The output of $Cls$, i.e. $\hat{Y_0}$, is used to evaluate the performance of this model.

\subsection{Baseline Methods}
To verify the performance of RAE, we used three DL models, i.e., DeepConvNet~\cite{DeepConvNet}, Deep Convolution neural network (DCNN)~\cite{app9142870}, EEGNet~\cite{Lawhern_2018}, as the compared baseline methods. The accuracy, sensitivity and specificity were served as the evaluation metrics of classification performance.

\subsubsection{DeepConvNet}
DeepConvNet is a deep convolution model proposed by Schirr-meister et al~\cite{DeepConvNet}. This model contains four convolution blocks. One temporal convolution filter and one max pooling layer are employed in each block. In particular, a spatial convolution layer is added additionally. Due to its robustness and high performance, DeepConvNet is widely used in the field of classification based on EEG~\cite{8607897,8709723}.

\subsubsection{DCNN}
Deep Convolution neural network (DCNN) is a method specifically used for SZ classification proposed by Oh et al~\cite{app9142870}. DCNN consists of five convolution layers, two max pooling layers, two average pooling layers, a global average pooling layer and a fully-connected layer. The convolution layers are able to extract features automatically, and the max pooling layers are able to capture the most significant feature extracted by the previous convolution layer.  Finally, all features are used to classify the signal in the fully-connected layer. DCNN achieved 81.26\% accuracy with the time window of length 25 s.

\subsubsection{EEGNet}
EEGNet is a compact convolutional neural network proposed by Lawhern et al~\cite{Lawhern_2018} . They first introduced the use of depthwise convolution and separable convolution on the EEG data. EEGNet also applied several well-known ideas in the field of BCI, such as optimal spatial filtering and filter bank. Due to the compact structure and stable performance, EEGNet has been widely applied in EEG-based classification tasks, such as steady-state visual evoked potential~\cite{Waytowich_2018}, motor imagery~\cite{10.3389/fnins.2019.01275}, and emotion recognition~\cite{8489715}, etc.

\section{Experiments and Results}

\subsection{Model Implementation}
​To evaluate the performance of the methods, level-one-subject-out (LOSO) strategy was used. Specifically, the data from one subject was used as the testing data, those of the remaining subjects were adopted as the training data. This procedure repeated until all the subjects served as the testing subject once. Each method was run for 5 times, and then the average accuracy, sensitivity and specificity were calculated as evaluation indicators via equations (\ref{eq_acc}) to (\ref{eq_spe}), which are usually used in the disease detection field~\cite{a14050139}.
\begin{equation}
	{accuracy = \dfrac{TP+TN}{TP+FP+TN+FN} } \label{eq_acc}
\end{equation}
\begin{equation}
	{sensitivity = \dfrac{TP}{TP+FN} } \label{eq_sen}
\end{equation}
\begin{equation}
	{specificity = \dfrac{TN}{TN+FP} } \label{eq_spe}
\end{equation}
where ​TP, TN, FP and FN denotes the total number of true positive, true negative, false positive and false negative examples, respectively.

For the RAE and all compared models, adaptive moment estimation (ADAM) optimizer was adopted as the optimization method~\cite{kingma2014adam}, and the learning rate was set as 1e-4. The experiment was executed for 30 iterations, and the accuracy in the last epoch was employed to evaluate the performance of all the methods.

\begin{table}[!htb]
	\centering
	\renewcommand{\arraystretch}{1.15}
	\setlength{\tabcolsep}{7pt}
	\caption{ The classification accuracies of each subject (mean$\pm$std.) for the RAE and all compared methods. }
\scalebox{0.9}{	\begin{tabular}{cllll}
		\hline
		\hline
		Sub.    &  DeepConvNet       &  DCNN       &  EEGNet       &  RAE        \\
		\hline
		1    &  42.47		$\pm$ 52.76       		&	95.21  	$\pm$ 3.96     		&	91.23	$\pm$ 6.75     		&	97.95	$\pm$ 3.28		\\
		2    &  44.58		$\pm$ 50.13       		&	15.52   	$\pm$ 6.24       	&	10.94	$\pm$ 11.53       	&	21.67	$\pm$ 6.4		\\
		3    &  81.38		$\pm$ 41.65       		&	100. 		$\pm$ 0.    			&	99.5		$\pm$ 0.52     		&	99.12	$\pm$ 1.63		\\
		4    &  59.75		$\pm$ 54.27     		&	53.74 	$\pm$ 31.61       	&	22.09	$\pm$ 41.6       	&	94.6		$\pm$ 7.81		\\
		5    &  84.84		$\pm$ 31.55       		&	95.05   	$\pm$ 2.95     		&	99.57	$\pm$ 0.7     		&	99.78	$\pm$ 0.48		\\
		6    &  64.64		$\pm$ 36.98       		&	24.9   	$\pm$ 15.08     	&	53.38	$\pm$ 15.03     	&	61.99	$\pm$ 12.73	\\
		7    &  53.29		$\pm$ 47.85     		&	49.76   	$\pm$ 22.53       	&	40.35	$\pm$ 14.29     	&	46.59	$\pm$ 4.79		\\
		8    &  69.51		$\pm$ 45.01       		&	82.68   	$\pm$ 36.34       	&	98.29	$\pm$ 1.69     		&	88.05	$\pm$ 11.13	\\
		9    &  60.			$\pm$ 54.77       		&	80.25   	$\pm$ 44.16       	&	100.		$\pm$ 0.     			&	99.88	$\pm$ 0.28		\\
		10  &   80. 		$\pm$ 44.72       		&	84.07   	$\pm$ 11.72       	&	96.91	$\pm$ 2.83     		&	97.28	$\pm$ 3.45		\\
		11  &   65.29 	$\pm$ 48.36       		&	77.53   	$\pm$ 35.11       	&	100.		$\pm$ 0. 				&	99.88	$\pm$ 0.26		\\
		12  &   42.73  	$\pm$ 47.2       		&	33.29   	$\pm$ 35.48       	&	37.27	$\pm$ 12.18   		&	24.6		$\pm$ 8.45		\\
		13  &   58.92  	$\pm$ 46.28       		&	98.44 	$\pm$ 2.14     		&	99.04	$\pm$ 0.91     		&	98.56	$\pm$ 1.24		\\
		14  &   61.56  	$\pm$ 52.71       		&	47.19   	$\pm$ 21.69       	&	32.93	$\pm$ 19.6       	&	62.16	$\pm$ 28.35	\\
		15  &   31.1  	$\pm$ 43.1         		&	53.05	$\pm$ 13.29   		&	35.37	$\pm$ 13.45      	&	36.46  	$\pm$ 8.39		\\
		16  &   79.75  	$\pm$ 44.58         	&	72.03  	$\pm$ 35.11   		&	93.67	$\pm$ 4.54      	&	86.33  	$\pm$ 20.86	\\
		17  &   40. 		$\pm$ 54.77         	&	78.77  	$\pm$ 9.58     		&	75.61	$\pm$ 7.51      	&	68.16  	$\pm$ 15.98	\\
		18  &   75.89  	$\pm$ 43.2         		&	74.25	$\pm$ 12.15   		&	99.73	$\pm$ 0.61      	&	97.53  	$\pm$ 1.79		\\
		19  &   20. 		$\pm$ 44.72         	&	90.26  	$\pm$ 7.21		 	&	88.55	$\pm$ 9.04      	&	98.95  	$\pm$ 1.2		\\
		20  &   39.69  	$\pm$ 39.57         	&	39.43  	$\pm$ 7.87    		&	51.38	$\pm$ 9.2      		&	76.48  	$\pm$ 4.19		\\
		21  &   60. 		$\pm$ 54.77         	&	96.58  	$\pm$ 7.36    		&	100.		$\pm$ 0.  			&	99.79  	$\pm$ 0.28		\\
		22  &   79.89  	$\pm$ 44.36         	&	92.28  	$\pm$ 8.07    		&	97.72	$\pm$ 1.41      	&	98.8  	$\pm$ 0.45		\\
		23  &   40. 		$\pm$ 54.77         	&	94.29  	$\pm$ 1.9    		&	95.86	$\pm$ 3.79  		&	92.02  	$\pm$ 7.99		\\
		24  &   93.78  	$\pm$ 13.91         	&	99.27  	$\pm$ 0.67    		&	97.44	$\pm$ 2.74      	&	98.17  	$\pm$ 3.42		\\
		25  &   60. 		$\pm$ 54.77         	&	30.  		$\pm$ 9.05      	&	51.52	$\pm$ 12.97     	&	55.22  	$\pm$ 16.47	\\
		26  &   40. 		$\pm$ 53.22       		&	93.93  	$\pm$ 2.57    		&	96.82	$\pm$ 1.43      	&	92.94  	$\pm$ 3.94		\\
		27  &   80. 		$\pm$ 44.72         	&	99.64  	$\pm$ 0.54    		&	99.88	$\pm$ 0.27  		&	99.28  	$\pm$ 1.3		\\
		28  &   41.84 	$\pm$ 53.23         	&	98.37  	$\pm$ 2.29    		&	96.12  	$\pm$ 6.76      	&	98.37  	$\pm$ 2.02		\\
		\hline
		\textbf{Mean} & 58.96$\pm$6.92 &  73.21$\pm$4.74  &  77.18$\pm$0.96  &  \textbf{81.81$\pm$1.60}\\
		\hline
		\hline
	\end{tabular}}
	\label{table1}
	
\end{table}

\subsection{Results}

The classification results of each subject obtained by all the methods are summarized in Table~\ref{table1}. For each subject, the average accuracy was calculated by averaging the accuracies of five times experiments, and the standard deviation was also calculated on these accuracies. We could find that the proposed RAE achieved better performance than all other methods, which yields the average accuracy of 81.81\%.  Besides, the results indicate that the RAE could yield more robust results with smaller standard deviations, such as those of subject No.4.

Since the intra-subject SEN and SPE have no significance owing to the unique label of data from each subject, we summarized the global confusion matrix on all the five experiments and calculated SEN and SPE across subjects. The results are listed in Table~\ref{table12}. ACC denotes the average accuracy across the subjects.
\begin{table}[!htb]
	\centering
	\renewcommand{\arraystretch}{1.15}
	\setlength{\tabcolsep}{7pt}
	\caption{Classification results of RAE and all compared methods. ACC, SEN and SPE denotes accuracy (mean$\pm$std), sensitivity and specificity, respectively.}
	\scalebox{0.85}{\begin{tabular}{lccc}
		\hline
		\hline
		Methods  & ACC(\%) & SEN(\%) & SPE(\%)  \\
		\hline
		DeepConvNet & 58.96$\pm$6.92 & 60.24 & 55.33\\
		DCNN       & 73.21$\pm$4.74  & 71.91 & 75.18  \\
		EEGNet       & 77.18$\pm$0.96  & 74.58 & 79.36  \\
		\textbf{RAE}       & \textbf{81.81$\pm$1.60}  & \textbf{80.30} & \textbf{83.37}  \\
		\hline
		\hline
	\end{tabular}}
	\label{table12}
\end{table}

\section{Discussion and Conclusion}
The results indicate that RAE is an effective method for SZ detection. It is worth mentioning that RAE could serve as a model framework, the detailed structure could be adjusted according to specific classification tasks. Namely, the backbone of the encoder can be adapted to the tasks, and the selection of backbone will greatly affect the performance of model. Besides, the number of loops ($n$) could be optimized according to the classification task. We conducted a series of experiments to obtain the best value of $n$, and each experiment was implemented five times. As shown in the Table~\ref{n_loop}, when $n$ was set to 2, the model obtained the best accuracy and relatively balanced sensitivity and specificity.

In the current study, only one dataset was used to evaluate the performance, more SZ datasets should be collected to verify the generalization of RAE. Besides, RAE is expected to be effective in detecting other mental diseases, such as major depressive disorder. We have conducted several preliminary experiments and will release the further results in the future studies.

\begin{table}[!htb]
	\centering
	\renewcommand{\arraystretch}{1.15}
	\setlength{\tabcolsep}{7pt}
	\caption{Result of experiments concerning $n$ selection.}
	\scalebox{0.95}{\begin{tabular}{lccc}
			\hline
			\hline
			$n$  & ACC(\%) & SEN(\%) & SPE(\%)  \\
			\hline
			1 & 77.74$\pm$2.03 & 75.55 & 79.13\\
			\textbf{2}       & \textbf{81.81$\pm$1.60}  & \textbf{80.30} & \textbf{83.37}  \\
			3 & 79.61$\pm$1.36 & 78.46 & 80.74\\
			4 & 78.26$\pm$2.72 & 76.08 & 82.13\\
			5 & 77.67$\pm$1.48 & 75.68 & 79.00\\
			\hline
			\hline
	\end{tabular}}
	\label{n_loop}
\end{table}

In summary, we proposed a novel framework method for SZ detection with recurrent Auto-Encoder. This method achieved an average accuracy of 81.81\%, sensitivity of 80.30\%, and specificity of 83.37\% in the LOSO experiments, which improved 4.62\% than the best baseline method. The RAE is expected to be a feasible tool in clinical diagnosis benefited by its superior performance and stability.

\section*{Acknowledgments}
This work was supported in part by the National Natural Science Foundation of China under Grant No.62076209.
%
%
%
 \bibliographystyle{splncs04}
 \bibliography{citelist}
%
%
%
%
%
\end{document}